\documentclass[%
 reprint,
 amsmath,amssymb,
 aps,
floatfix,
]{revtex4-2}
\usepackage[normalem]{ulem}
\usepackage{graphicx}
\usepackage{dcolumn}
\usepackage{bm}
\usepackage{xcolor}

\begin{document}

\preprint{APS/123-QED}

\title{A classical model for sub-Planckian thermal diffusivity in complex crystals }

\author{Huan-Kuang Wu}
\affiliation{Department of Physics, Condensed Matter Theory Center and Joint Quantum Institute, University of Maryland, College Park, MD 20742, USA}

\author{Jay D. Sau}
\affiliation{Department of Physics, Condensed Matter Theory Center and Joint Quantum Institute, University of Maryland, College Park, MD 20742, USA}

\date{\today}

\begin{abstract}

Measurements of thermal diffusivity in several insulators have been shown to 
reach a Planckian bound on thermal transport that can be thought of as the limit of validity of semiclassical phonon scattering. Beyond this regime, the heat transport must be understood in terms of incoherent motion of the atoms under strongly anharmonic interactions. In this work, we propose a model for heat transport in a strongly anharmonic system where the thermal diffusivity can be lower than the Planckian thermal diffusivity bound. Similar to the materials which exhibits thermal diffusivity close to this bound, our scenario involves complex unit cell with incoherent intra-cell dynamics. We derive a general formalism to compute thermal conductivity in such cases with anharmonic intra-cell dynamics coupled to nearly harmonic inter-cell coupling. Through direct numerical simulation of the non-linear unit cell motion, we explicitly show that our model allows sub-Planckian thermal diffusivity. We find that 
the propagator of the acoustic phonons becomes incoherent throughout most of the Brillouin zone in this limit. We expect these features to apply to more realistic models of complex insulators showing sub-Planckian thermal diffusivity, suggesting a multi-species generalization of the thermal diffusivity bound that is similar to the viscosity bound in fluids.
\end{abstract}

\maketitle


\section{Introduction}

Despite the diversity of strongly interacting quantum materials, the low energy response of such systems is mostly found to be described by weakly interacting excitations. Paradigmatic examples of these are the quasiparticles in Fermi liquid theory or Goldstone bosons such as phonons or magnons~\cite{landau2013course}. Notable exceptions to this, in the form of so-called incoherent metals~\cite{bruin2013similarity,zaanen2004temperature}, have been proposed in the form of non-Fermi liquids~\cite{hertz1976quantum,stewart2001non,lohneysen2007fermi,lee2018recent} and Bose metals~\cite{kapitulnik2019colloquium}, though definitive experimental evidence for such phases is lacking. 
Signature of incoherent metals were discovered experimentally in high-Tc superconductors~\cite{takagi1992systematic,hussey1997anisotropic,ando2002electrical,takenaka2003incoherent,bach2011high}, where linear-$T$ resistance appears and persists beyond the Mott-Ioffe-Regel limit~\cite{emery1995superconductivity} where the mean free path becomes smaller than the Fermi wavelength.
Surprisingly, the linear-$T$ resistance is also observed at apparently low temperatures~\cite{bruin2013similarity}.  Such a scaling is in direct contradiction to the $T^2$ behavior expected from the low temperature limit of Boltzmann scattering transport of Fermi liquid quasiparticles~\cite{ziman2001electrons} from electron-electron interaction.  On the other hand, if the electrons are assumed to scatter from other low-energy excitations with a scattering rate $\tau_s^{-1}$, then the conductivity of a material in the semiclassical approximation is expected to follow the Drude formula $\sigma\propto \omega_p^2\tau_s$ \cite{kittel1996introduction, zaanen2004temperature}, where $\omega_p$ is the plasma frequency. The existence of well-defined quasiparticles, which are the crucial ingredients for Boltzmann theory, are expected to be meaningful when their lifetime induced energy broadening $\hbar\tau^{-1}$ is smaller than the energy of the quasiparticles. Since the typical energy of quasiparticles are of order $k_B T$, this leads to the suggestion of a minimal conductivity $\sigma>\omega_p^2 \tau_P$ where $\tau_P=k_B T/\hbar$ is the Planck scattering time~\cite{zaanen2004temperature}. 
Interestingly, a significant number of systems presents conductivity near the minimal limit~\cite{bruin2013similarity,homes2004universal,legros2019universal,nakajima2020quantum}. 
The description of transport in such systems requires a fully quantum mechanical treatment, which has inspired significant theoretical effort~\cite{shaginyan2013quasiclassical,hartnoll2015theory,shaginyan2019fermion,patel2019theory,volovik2019flat}.

These ideas have been extended~\cite{hartnoll2015theory} beyond the semiclassical regime to relate more general transport coefficients such as momentum, charge and heat diffusion to viscosity bounds that had been proposed based on holographic methods~\cite{kovtun2005viscosity}. 
A similar instance of universal diffusion bound described by fundamental physical constants is also discovered in liquid systems~\cite{trachenko2020minimal,baggioli2020similarity,trachenko2021universal}.
In fact, the heat or thermal diffusion coefficient in incoherent metals, where quasiparticles are absent,
was studied extensively~\cite{blake2016universal,blake2017thermal,gu2017local,davison2017thermoelectric,patel2017quantum,blake2017diffusion} and 
was shown to be related to the scrambling time~\cite{maldacena2016bound}, which is be bounded by $\lambda_L^{-1} \geq 2\pi k_BT/\hbar$ whose form is close to the Planckian time $\tau_P$. 
 A complication of measuring thermal diffusion, as was done in cuprate superconductors in the bad metal regime~\cite{zhang2017anomalous,zhang2019thermal}  is that it contains contributions from both phonons and electrons.
 The thermal diffusivity contributions from electrons and phonons in materials approaching the Planckian bound are expected to be described by the form  
\begin{align}
D_P = v_s^2\tau_P
\label{Eq:Planckian_diffusivity}
\end{align}
where $v_s$ is the sound velocity or the Fermi velocity depending on the relevant carrier \cite{zhang2019thermal}. 
Accordingly, we restrict our discussion to crystalline systems with a well-defined sound velocity in this work.
Correlating the thermal diffusion and charge diffusion measurements leads to the conclusion that 
 that the electron and phonon behaves like a soup where both contribute to the thermal transport in an incoherent way~\cite{zhang2019thermal}. 
A simpler testing ground for these ideas are provided by the thermal diffusion in insulators where the thermal diffusion 
is contributed exclusively by phonons. 
In this case it has been proposed~\cite{martelli2018thermal,behnia2019lower} that Eq.~\ref{Eq:Planckian_diffusivity} provides a lower bound for thermal diffusivity for temperatures at or above the order of Debye temperature $T_D$.
Interestingly, this bound appears to be reached for insulators with complex unit cell~\cite{martelli2018thermal,behnia2019lower}. Such slow thermal diffusivity has been attributed to be a signature of quantum chaos~\cite{zhang2019thermalization}. 

Recently, Mousatov et al.~\cite{mousatov2020planckian} have pointed out that the diffusivity bound (Eq.~\ref{Eq:Planckian_diffusivity}) can be understood to be a consequence of the fact that the sound velocity $v_s$ is bounded by the melting velocity $v_M$. This suggests that the thermal diffusivity in complex oxides where $v_s$ approaches $v_M$ could approach the bound in Eq.~\ref{Eq:Planckian_diffusivity}.
It is also possible that this thermal diffusivity bound motivated by the connection~\cite{hartnoll2015theory} to the viscosity bound is modified for complex insulators. This is because the viscosity bound has been shown to be lowered in multi-component fluids~\cite{kovtun2005viscosity}. 

In this paper, we study the thermal diffusion in a model of a strongly anharmonic crystal and show that in certain parameter regimes the thermal diffusivity can drop 
below the Planckian bound given in Eq. \ref{Eq:Planckian_diffusivity}. To simplify the problem, we will assume 
that the temperature $T$ of the system is high enough that the dynamics of the atoms in the crystal can be approximated as classical. 
However, Planck's constant $\hbar$ enters through the requirement that all phonon frequencies must be below $k_BT/\hbar$. The model we consider has a unit cell with a large number of atoms with very anharmonic interactions that lead to incoherent intracell atomic dynamics. 
We will show that these modes contribute negligibly to heat transport, while contributing to heat capacity in a way similar to Einstein phonons~\cite{einstein1911elementary}. 
This reduces the thermal diffusivity of the system.  
In Sec. \ref{sec:2}, we will first discuss the complex phonon system in the context of Boltzmann transport theory and show that thermal diffusivity must obey the Planckian bound (Eq.~\ref{Eq:Planckian_diffusivity}) as long as all phonons are well-defined in the system. 
In Sec. \ref{sec:3}, we derive an expression for thermal diffusivity of lattice systems with strongly anharmonic intra-cell dynamics connected by weakly anharmonic springs. In Secs. \ref{sec:4} and \ref{sec:5}, we construct and simulate an example of the model discussed in Sec. \ref{sec:3} and demonstrate the breaking of the Planckian thermal diffusivity. 

\section{Thermal Diffusion in the Boltzmann Regime}
\label{sec:2}
The thermal transport of a complex phonon system in the classical regime is qualitatively captured by the Boltzmann transport theory in many cases. Under the relaxation time approximation, the thermal conductivity $\kappa$, as derived by Peierls~\cite{peierls1929kinetischen}, takes a form similar to that from kinetic theory,
\begin{align}
    \kappa = \frac{1}{d}\sum_{\vec{q},\lambda}C(\vec{q},\lambda)v^2(\vec{q},\lambda)\tau(\vec{q},\lambda)
    \label{Eq:drude}
\end{align}
where $d$ is the dimension, $\vec{q}$ and $\lambda$ are the wave vector and mode index, respectively. $C$ is the specific heat per unit volume, $v$ is the mode velocity, and $\tau$ is the relaxation time. In the classical regime, all normal modes satisfy equipartition principle. That is, $T > \hbar\omega_{max}$ where $\omega_{max}$ represents the highest phonon frequency in the system, and $C(\vec{q},\alpha)$ can be approximated by $k_B/V$. In this case, the thermal diffusivity $D_{th} \equiv \kappa/\sum_{\vec{q},\lambda}C(\vec{q},\lambda)$ is given by
\begin{align}
    D_{th} =\frac{1}{Nd} \sum_{\vec{q},\lambda}v^2(\vec{q},\lambda)\tau(\vec{q},\lambda),
    \label{Eq:thermal_diffusivity}
\end{align}
where $N$ represents the number of modes. 

The central assumption behind the Boltzmann formalism requires the phonons to be well-defined. Therefore, the scattering rate $\tau^{-1}$ should not exceed the frequency spacing between different modes. Assuming equal level spacing, this further imposes a lower bound on $\tau(\vec{q},\lambda)$,
\begin{align}
    \tau(\vec{q},\lambda) > \frac{N}{\omega_{max}(\vec{q})}>N\tau_P,
    \label{Eq:3}
\end{align}
where the second inequality comes from the classical requirement with $\omega_{max}(\vec{q})$ being the highest optical phonon frequency with momentum $\vec{q}$. That is, the distribution is classical equipartition rather than Bose-Einstein distribution. $\tau_P \equiv \hbar/k_B T$ is the Planckian time. Plugging in Eq. \ref{Eq:3} to Eq. \ref{Eq:drude} and taking the fastest phonon velocity to be the longitudinal sound velocity $v_s$, we can deduce a lower bound for thermal diffusivity within the Boltzmann regime,
\begin{align}
    D_{th} > \frac{1}{d}v_s^2\tau_P\sim D_P.
\end{align}

Therefore, the energy diffusion obtained from Boltzmann transport of phonon will be bounded by the Planckian diffusivity~\cite{hartnoll2015theory,zhang2017anomalous,behnia2019lower}. 
{}

\section{Thermal transport with incoherent intra-cell dynamics}
\label{sec:3}
To go beyond Boltzmann regime, we consider a lattice model with highly non-linear intra-cell dynamics as illustrated in Fig. \ref{Fig.1}(a). The unit cells are coupled to each other through weakly anharmonic springs acting on an external degrees of freedom $\vec{r}=(r_x,r_y,r_z)$ to form a 3d lattice. Such a model can be considered as a more tractable version of atomic motion in insulators with complex unit cell where thermal diffusivity close to the Planckian limit has been reported~\cite{zhang2019thermalization}, which will be elaborated in Sec. \ref{sec:6}. 

The spring force on the $i^{th}$ unit cell (circles in Fig. \ref{Fig.1}(a)) is written as
\begin{align}
    \vec{F}(i) = \sum_{\alpha\in\{x,y,z\}} \vec{f}(i,\alpha)-\vec{f}(i-\alpha,\alpha)
    \label{eq:eom1}
\end{align}
where $\vec{f}(i,\alpha)$ represents the force from the spring to the $\alpha$ direction of site $i$
\begin{align}
    \vec{f}(i,\alpha) = -k(\vec{r}_i - \vec{r}_{i+\alpha})+A\sum_{u\in\{x,y,z\}}\left({r}_{u,i} - r_{u,i+\alpha}\right)^2\hat{u}
    \label{eq:spring_force}
\end{align}
with spring constant $k$ and weak cubic anharmonicity $A$. Next, 
at finite temperature $T$, we assume the intra-cell dynamics to be described by the response function $\chi$. That is, 
\begin{align}
    \vec{r}_i(t) = \int dt' \chi(t-t')\vec{f}_i(t')
    \label{eq:eom2}
\end{align}
where $\vec{f}_i$ represents a force acting on $\vec{r}_i$ and $\chi(t)=0$ for all $t<0$. Note that since the center of mass motion of the unit cell is decoupled from any intra-cell motion, the acoustic phonons at long wavelength will not be damped efficiently by the intra-cell dynamics. This is the reason for having to introduce $A$ in Eq. \ref{eq:spring_force}, which in turn leads to damping of the acoustic modes. In principle, Eqs. \ref{eq:eom1}, \ref{eq:spring_force}, and \ref{eq:eom2} constitute the equations of motion for the unit cells and one can determine the trajectories $\vec{r}_i(t)$ by solving them self-consistently.

As mentioned in Sec. \ref{sec:2}, the Boltzmann formalism does not apply in regimes with ill-defined phonons such as the case of strongly anharmonic intra-cell interactions. In this case, one can instead use the Green-Kubo formula~\cite{zwanzig1965time} given by 
\begin{align}
    \kappa = \frac{1}{k_BT^2}\int_0^\infty d\tau\sum_n\left\langle q_{i,\alpha}(t)q_{i+n,\alpha}(t+\tau)\right\rangle.
    \label{eq:GK}
\end{align}
where $q_{i,\alpha}$ represents the heat flux in the $\alpha$ direction at site $i$ which can be written as the rate of energy transfer across the spring connecting sites $i$ and $i+\alpha$,
\begin{align}
    q_{i,\alpha} = -\frac{\vec{f}(i,\alpha)\cdot\left(\dot{\vec{r}}_i+\dot{\vec{r}}_{i+\alpha}\right)}{2}.
    \label{eq:heat_flux}
\end{align}
A crucial assumption for the consistency of Eq. \ref{eq:heat_flux} is that the average power absorbed by the spring
\begin{align}
    \tilde{q}_{i,\alpha} = -\vec{f}(i,\alpha)\cdot\left(\dot{\vec{r}}_i-\dot{\vec{r}}_{i+\alpha}\right)
\end{align}
vanishes. This is clearly satisfied by the conservative force in Eq. \ref{eq:spring_force}.

Ultimately, the combination of Eqs. \ref{eq:eom1}, \ref{eq:spring_force}, and \ref{eq:eom2} is a complex non-linear system of equations that requires numerical molecular dynamics to solve~\cite{ladd1986MD,hoover2012computational}. However, in the limit of weak anharmonicity, the anharmonic part of the force on a spring i.e.  $\vec{f}^A(i,\alpha)=A\sum_{u}\left(r_{u,i}-r_{u,i+\alpha}\right)^2\hat{u}$ can be approximated as being random and uncorrelated $\vec{f}^A(i,\alpha)\approx \vec{\eta}(i,\alpha)$. The mean value of $\vec{\eta}(i,\alpha,t)$ contributes to thermal expansion~\cite{Ashcroft} and can be set to zero by shifting the lattice constants. The variance of $\vec{\eta}$ will be determined self-consistently by solving the combination of Eqs. \ref{eq:eom1}, \ref{eq:spring_force}, and \ref{eq:eom2} at finite $T$ (see App. \ref{appendix:B}). A random stochastic driving force in a spring would violate the vanishing of the average power $\tilde{q}$ absorbed by the spring. This is remedied by adding a damping term with a coefficient $\lambda$ determined by the fluctuation-dissipation theorem as
\begin{align}
    \left\langle\eta_u(i,\alpha,t)\eta_v(j,\beta,t')\right\rangle = 2\lambda k_B T\delta_{u,v}\delta_{i,j} \delta_{\alpha,\beta}\delta(t-t').
    \label{eq:force_correlation}
\end{align}
where $u,v\in \{x,y,z\}$ represents the different components of the $\vec{\eta}$. The anharmonic part of the force approximated as the combination of random and damping terms is written as:
\begin{align}
    \vec{f}^A(i,\alpha)\approx -\lambda (\dot{\vec{r}}_i - \dot{\vec{r}}_{i+\alpha})+\vec{\eta}(i,\alpha,t).
    \label{eq:linearized_force}
\end{align}
The above equation together with Eqs. \ref{eq:spring_force} and \ref{eq:eom2} are now linear so that the correlation functions of the position are Gaussian. The distribution is then completely determined by the 2-point correlation function $\langle r_{u,i}(0)r_{v,j}(t)\rangle$, which can be found by solving Eq. \ref{eq:eom1}, \ref{eq:spring_force}, \ref{eq:eom2}, \ref{eq:force_correlation}, and \ref{eq:linearized_force} in a self-consistent way.

Within the Gaussian approximation for the position distributions, the higher order correlation functions in Eq. \ref{eq:GK} can be expanded, using Wick's theorem, into products of 2-point correlation functions, or particularly, position-position power spectrum, $S_{xx}(\vec{q},\omega)\equiv \sum_j\int d\tau e^{-i(\vec{q}\cdot \vec{r}_{ij}-\omega\tau)}\left\langle r_{x,i}(t)r_{x,j}(t+\tau)\right\rangle$ (See App. \ref{appendix:A} for details). According to fluctuation-dissipation theorem~\cite{kubo1966fluctuation}, $S_{xx}(\vec{q},\omega)$ is related to the imaginary part of the response function by
\begin{align}
    S_{xx}(\vec{q},\omega) = \frac{2k_B T}{\omega} \text{Im}\left[D(\vec{q},\omega)\right]
    \label{Eq:power_spectrum}
\end{align}
where $D(\vec{q},\omega)$ is the response function in frequency-momentum space defined by
\begin{align}
    \vec{r}(\vec{q},\omega) = D(\vec{q},\omega)\vec{F}(\vec{q},\omega).
\end{align}

As shown in App. \ref{appendix:B}, the response function $D$ can be approximated by that of a damped phonon system which is written as:
\begin{align}
    D(\vec{q},\omega) = \frac{1}{\chi^{-1}(\omega)+4(k-i\omega\lambda)\sum_{u\in\{x,y,z\}}\sin^2 q_u/2}.
    \label{eq:phonon_GF}
\end{align}
The damping coefficient $\lambda$ , which scales linearly in $T$ (as derived in the appendix), is related to the anharmonic force through fluctuation-dissipation theorem~\cite{kubo1966fluctuation}.

\section{Shell-ball Model for Unit Cell}
\label{sec:4}
\begin{figure}
\centering
\includegraphics[width=0.46 \textwidth]{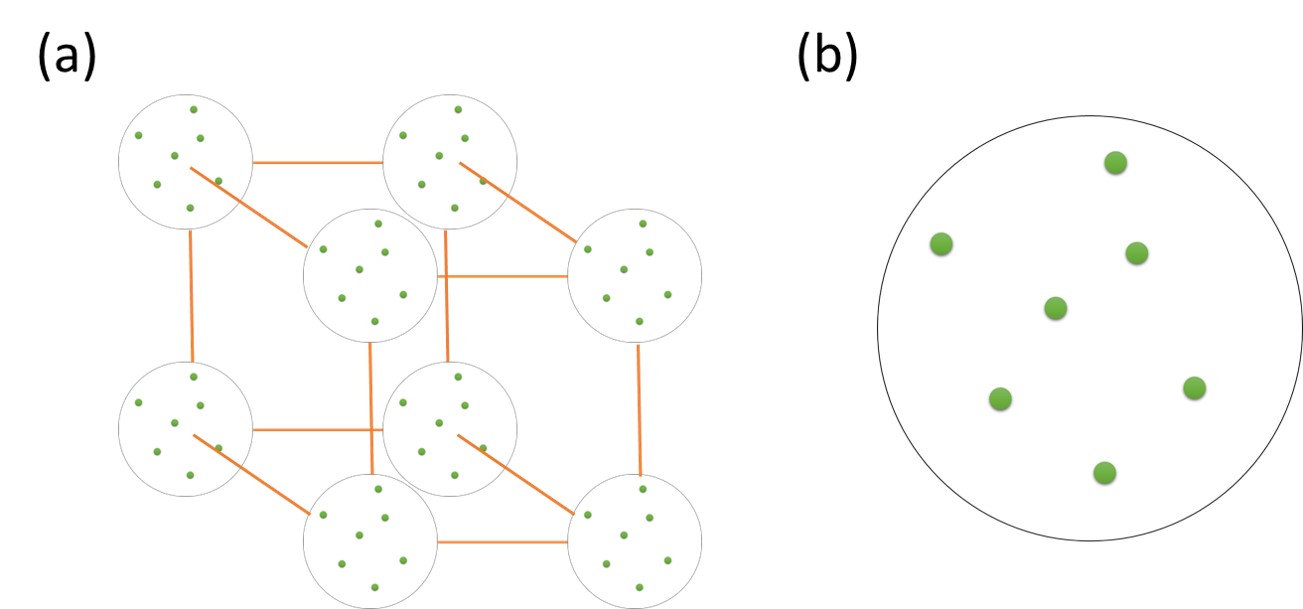}
\caption{(a)Schematics of our cubic lattice. The orange lines represent non-linear springs. Each unit cell is composed of free balls (green circles) of mass $m$ moving within a finite mass $M$ spherical shell of radius $R$, as illustrated in (b). }
\label{Fig.1}
\end{figure}

The response function $\chi$ in Eq. \ref{eq:eom2} in the last section is determined by the structure of the complex unit cell in Fig. \ref{Fig.1}. In this section, we consider a specific model for the complex unit cell consisting of $N$ identical balls with mass $m$ contained in a spherical shell of radius $R$ with mass $M$ as illustrated in Fig. \ref{Fig.1}. Within each shell, the balls act as point masses that do not interact with each other and move freely until colliding to the shell, which we assumed to be elastic. {\iffalse\color{red}At finite $M$, the phase space trajectory of the shell-ball system will be chaotic and satisfies equipartition principle among all degrees of freedom. As a result, the specific heat for a unit cell is $c_v = 3Nk_B/2$.\fi}
\begin{figure}
\centering
\includegraphics[width=0.48 \textwidth]{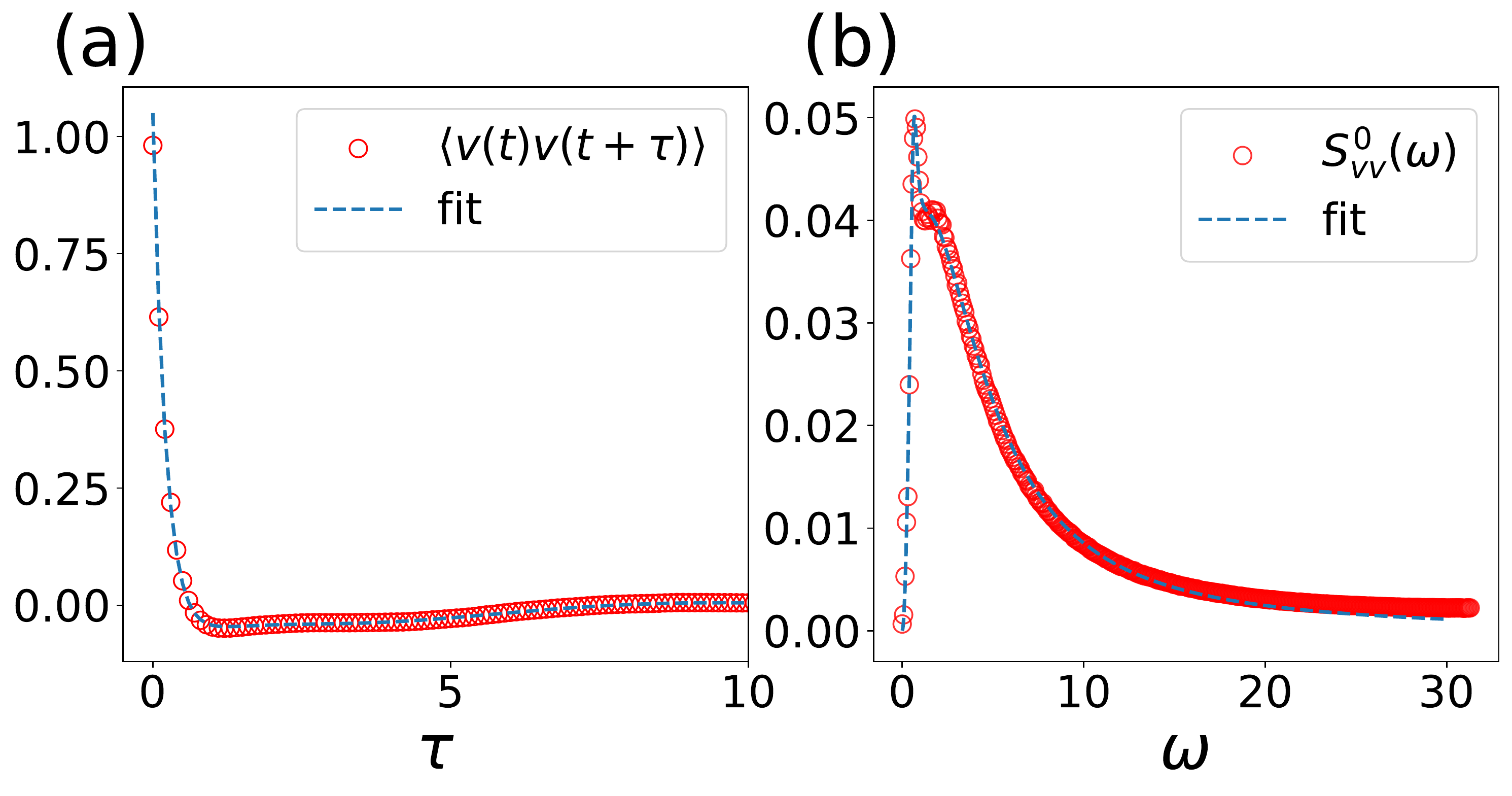}
\caption{(a) The velocity auto-correlation function for $N=40$ from simulation with $5\times 10^6$ collisions (red open circle). Blue dashed line is the inverse Fourier transform of the fit in the momentum space, as shown in (b). (b) Velocity-velocity power spectrum (red open circle) and  rational fit to the data (blue dashed line).}
\label{Fig.2}
\end{figure}

The shell-ball unit cell is studied by direct simulation. $N$ point masses with a total kinetic energy of $3Nk_BT/2$ are placed randomly inside a spherical shell of radius $R$. The shell velocity is inferred by $\vec{v}_{com}=0$. Next, we allow the balls to collide with the shell following energy and momentum conservation between the two. After each collision, the updated shell velocity  as well as the time of collision are recorded. These data can then be used to calculate the velocity probability distribution and velocity auto-correlation function. The total number of collisions is up to $5\times 10^6$ to reach statistical equilibrium after a warmup of $10^5$ collisions.

In this work, we will choose units so that $k_BT = 1$ and $m = 1$. The radius $R$ is chosen to be $10$ which is much larger than the thermal de Broglie wavelength $\lambda_{th} = \sqrt{2\pi/mk_BT}$ and the shell mass $M$ is chosen to be $M=m=1$. Fig. \ref{Fig.2} is the simulation result for $N = 40$ balls. The red open circles in Fig. \ref{Fig.2}(a) show the average over the velocity auto-correlation functions of the shell along $x$, $y$, and $z$ relative to the center of mass. By taking the Fourier transform, we obtain the velocity-velocity power spectrum as shown in Fig. \ref{Fig.2}(b). As expected, there is only one broad peak which locates at around the collision frequency of a single ball with the shell. This suggests that the intra-cell motion is mostly incoherent. To get a functional form for the power spectrum, we perform a rational-function fit on the data in Fig. \ref{Fig.2}(b). The resulting fit and its inverse Fourier transform are plotted as blue dashed line in Fig. \ref{Fig.2}(b) and Fig. \ref{Fig.2}(a), respectively.

\begin{figure}
    \centering
    \includegraphics[width=0.35\textwidth]{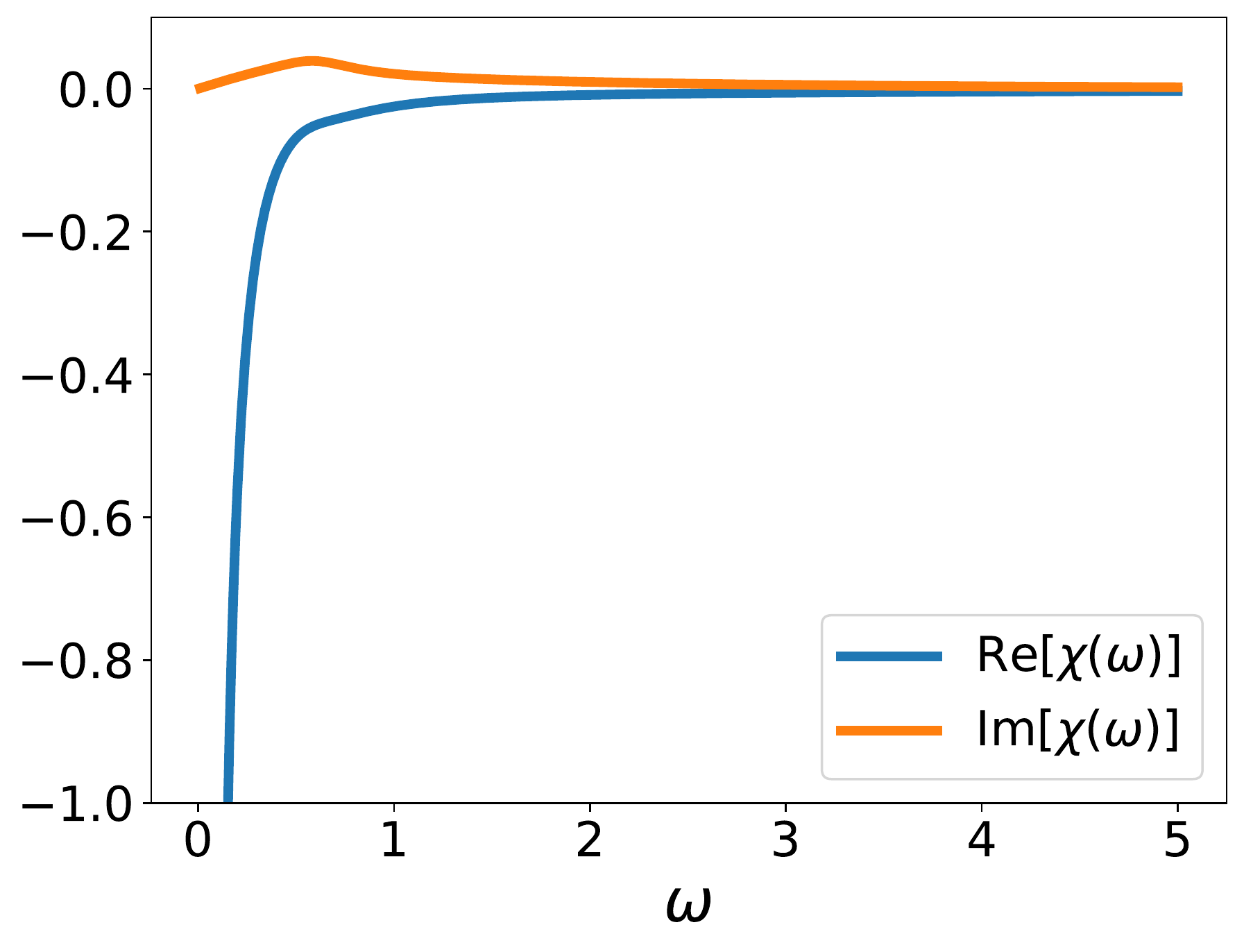}
    \caption{Real and imaginary parts of the response function $\chi(\omega)$ of the shell coordinate to an external force for $N=40$ balls.}
    \label{fig:chi}
\end{figure}

As shown in App. \ref{appendix:C}, the velocity distribution of the shell is Gaussian. Therefore, we expect the response of the shell coordinate $\vec{r}$ to external forces to be linear, consistent with Eq. \ref{eq:eom2}. Using fluctuation-dissipation theorem on $S^0_{xx}(\omega)=S^0_{vv}(\omega)/\omega^2$ and the Kramer-Kronig relation, we can obtain the imaginary and real parts of the response function $\chi_r(\omega)$ of the relative coordinate $\vec{r}_{shell}-\vec{r}_{com}$,
\begin{align}
    \text{Im}\left[\chi_r(\omega)\right] &= \frac{\omega}{2k_BT} S^0_{xx}(\omega) \nonumber\\
    \text{Re}\left[\chi_r(\omega)\right] &= \frac{1}{\pi}\int_{-\infty}^\infty d\omega'\frac{\omega'\text{Im}\left[\chi_r(\omega)\right]}{\omega'^{2}-\omega^2}.
\end{align}
Taking the center of mass motion into account, the response function of the shell coordinate to external force is given by
\begin{align}
    \chi(\omega) = -\frac{1}{(N+1)\omega^2}+\frac{N}{N+1}\chi_{r}(\omega).
\end{align}
The real and imaginary parts for $\chi$ with $N=40$ balls are shown in Fig. \ref{fig:chi} where the divergence at $\omega\rightarrow 0$ comes from the contribution of the center of mass motion.

\section{Thermal Diffusivity of the Shell-ball Model}
\label{sec:5}
\subsection{Acoustic Phonons in the Shell-ball Model}

In the limit of the highly overdamped complex shell-ball model discussed in Sec. \ref{sec:4}, the heat transport turns out to be dominated by acoustic phonons. From the Boltzmann transport equation Eq. \ref{Eq:drude}, the transport properties of the acoustic phonons are determined by their dispersion $\omega(\vec{q})$ and damping rate $\tau^{-1}(\vec{q})$. In the limit of weak damping, the acoustic mode is described by a phonon dispersion relation
\begin{align}
    \omega(\vec{q}) =\sqrt{\frac{4k\sum_u\sin^2q_u/2}{N+1}}
    \label{eq:freq}
\end{align}
with a broadening or inverse lifetime
\begin{align}
    \tau^{-1}(\vec{q})=\frac{2\lambda\sum_u\sin^2q_u/2}{N+1}.
    \label{eq:lifetime}
\end{align}

Note that the damping is quadratic in the long-wavelength limit, which is consistent with the Akhiezer’s damping~\cite{akhieser1939absorption}. However, microscopically, we considered the acoustic phonon to be underdamped, or $\omega\tau > 1$. This contrasts the regime for Akhiezer's mechanism ($\omega\tau<<1$), which is needed for the viewpoint of static lattice distortion at the timescale of relaxation. The choice of $k$ and $\lambda$ is restricted by the assumptions of our framework. That is, the acoustic phonon cannot be overdamped by $\lambda$, or $\tau^{-1}(\vec{q})<\omega(\vec{q})$, and the acoustic phonon frequency cannot exceed $k_BT = 1$. By substituting Eqs. \ref{eq:freq} and \ref{eq:lifetime} to the constraints above, we get the following conditions for the spring constant $k$ and damping constant $\lambda$:
\begin{align}
    \frac{6\lambda}{N+1}<\sqrt{\frac{12k}{N+1}}<1,
\end{align}
where the extreme case of $\vec{q}=(\pi,\pi,\pi)$ is taken.

For our numerical computation, we choose a set of parameters consistent with the constraint i.e. $k=2$ and $\lambda = 2$ for $N=40$ balls used in the calculation of chi in Fig. \ref{fig:chi}. In Fig. \ref{Fig.4}(a), we plot the imaginary part of the phonon Green's function $D(\vec{q},\omega)$ along $\vec{q}\parallel(1,1,1)$ in frequency-momentum space. As can be observed in bright color, the phonon Green's function exhibits a single coherent mode, while all the other degrees of freedom are incoherent. A vertical cut along $q=0.8$, as indicated by the red dashed line, is shown in \ref{Fig.4}(b). Besides the coherent Lorentzian peak, we can observe a broad background contributed by the incoherent modes around frequencies close to the peak in Fig. \ref{Fig.2}(b). Next, the real and imaginary part of the corresponding poles, denoted by $\omega_s$, is shown in \ref{Fig.4}(c) and \ref{Fig.4}(d). First, we confirm that the frequency and lifetime is in the desired regime by satisfying $\left|\text{Im}[\omega_s]\right|<\text{Re}[\omega_s]<1$. Next, they fit to the analytic form in Eqs. \ref{eq:freq} and \ref{eq:lifetime} (orange dashed line) at long wavelength, where the motion is mostly in-phase. This further confirms that this mode possesses the properties of the sound mode. In fact, the coherence of sound mode at long wavelength is expected due to the translational symmetry in our system.

\subsection{Sub-Planckian Thermal Diffusivity}

We are now ready to compute the thermal diffusivity of the shell-ball model. By plugging in the position-position power spectrum $S_{xx}(\vec{q},\omega)=(2k_BT/\omega)\text{Im}[D(\vec{q},\omega)]$ into Eqs. \ref{eq:GK} and \ref{eq:heat_flux}, we can obtain the thermal conductivity $\kappa$. Next, according to equipartition principle, the specific heat per unit cell is $3Nk_B/2$. This gives the thermal diffusivity of our system $D_{th} = 2\kappa/3Nk_B$. On the other hand, since a coherent sound mode exists, the Planckian thermal diffusivity $D_P\equiv v_s^2\tau_P$ is well-defined.
\begin{align}
    D_P = \frac{k}{N+1}.
\end{align}
For the parameter set considered here ($N=40$, $k=\lambda = 2$), the resulting $D_{th} \approx 8.13\times10^{-3}$, which is below the Planckian bound $D_P\approx4.88\times10^{-2}$. As a result, we have demonstrated a system with sub-Planckian thermal diffusion. 

\begin{figure}
\centering
\includegraphics[width=0.46 \textwidth]{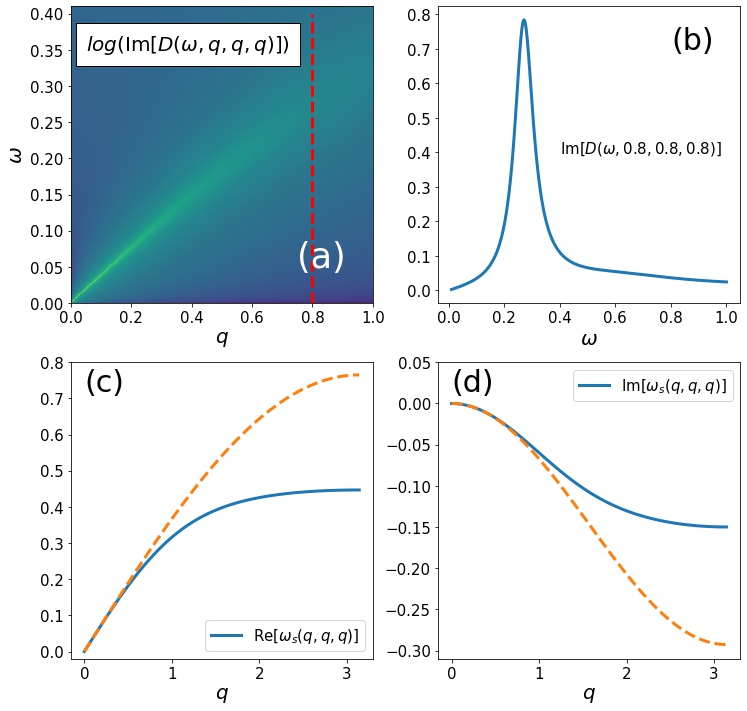}
\caption{(a) Logarithm of the imaginary part (color) of the phonon Green's function in the frequency-momentum space for $\vec{q}\parallel(1,1,1)$, where a coherent mode is visible. (b) The vertical cut of (a) along $q = 0.8$. (c) and (d) Blue solid lines: Real and imaginary part of the poles $\omega_s(q,q,q)$ along $q\parallel (1,1,1)$ that corresponds to the peaks in (a). Orange dashed lines: The corresponding analytic form from Eqs. \ref{eq:freq} and \ref{eq:lifetime}.}
\label{Fig.4}
\end{figure}

The mechanism for breaking the Planckian bound here is quite simple. Thermal diffusivity is defined as $D_{th}=\kappa/c_v$. For an $N$-degree-of-freedom unit cell, the specific heat per unit volume scales with $N$. 
However, since the phonon Green's function shows only one coherent peak (See Fig. \ref{Fig.4}(b)) corresponding to the acoustic mode, the optical phonons are incoherent. This is a direct consequence of the highly non-linear intracell dynamics. Due to the small relaxation time, these optical modes does not contribute significantly to the thermal conductivity and we expect the majority of the heat current to be carried by acoustic phonons.
In this case, the thermal conductivity $\kappa$ is only related to $N$ implicitly through the sound velocity $\kappa\sim v_s(N)^2$. As a result, the scaling of thermal diffusivity with $N$ is roughly $D_{th}\sim v_s(N)^2/N$. Therefore, we expect such scenario could attain $D_{th}/D_P$ that scales with inverse the number of balls and sub-Planckian thermal diffusion will appear in the large-$N$ regime.

\section{Discussion and Conclusions}
\label{sec:6}
The shell-ball model discussed here may be instructive for understanding the experimental results on Planckian thermal diffusion in materials with complex unit cells with a large number of atoms. In Ref.~\cite{zhang2019thermalization}, it has been pointed out that the insulators which presents thermal diffusivity close to $D_P$, such as the perovskites, usually exhibit complex unit cells. On the other hand, the insulators with simple unit cell usually have much larger value for $D_{th}/D_P$. This observation is consistent with our model, where the heat diffusivity is suppressed by the large $c_v$ from the large degrees of freedom $N$ within each unit cell, while the heat conductivity $\kappa$ does not scale linearly with $N$. Our model, unlike most holographic models, presents a low energy spectrum described by the coherent acoustic phonons. At low temperature, the thermal energy is carried by these excitations, leading to efficient heat diffusivity well above the Planckian bound. However, as we show, higher temperatures spread out the energy to higher frequency incoherent modes that are not efficiently transmitted, which results in suppressed heat diffusion below the Planckian bound. At intermediate temperature where some of the optical phonons are not thermally activated, we expect quantum behaviors to appear. Similar mechanism has been suggested in Ref.  \cite{zhang2019thermalization} to account for the appearance of $\hbar$ in the thermal diffusivity above Debye temperature. In addition, the intra-cell dynamics in our model is chaotic by nature due to strong non-linearity. This can also be connected to the proposal raised in Ref.  \cite{zhang2019thermalization} that the optical phonons in the Planckian materials likely exhibit chaotic dynamics. Finally, the scenario presented above suggests a smaller bound for thermal diffusivity that is roughly $D_P/N$. This behavior is consistent with the suppression of the viscosity bound in multi-component fluids~\cite{kovtun2005viscosity}. Given the degrees of freedom in the unit cell of the Planckian materials, this is only within one order of magnitude to the measured $D_{th}$. Therefore, we expect our discussion in Sec. \ref{sec:2} to be a useful aspect to explain the Planckian diffusion in experiments.

Recently, Ref. \cite{mousatov2020planckian} has suggested a diffusion bound based on the melting temperature. Specifically, the melting temperature $T_M$ would give rise to a velocity upper bound $v_M$ which then forces a lower bound to the phonon lifetime by using $l/v_M$, where $l$ is the phonon mean free path. If we introduce a melting temperature $T_M$ to our shell-ball lattice, a bound on the characteristic frequency $\omega_0=\sqrt{k/N+1}<T_M$ will appear so that the energy scale from quantum fluctuation on the springs does not cause the lattice to melt. The appearance of this bound can affect the Planckian diffusion bound (Eq. \ref{Eq:Planckian_diffusivity}) in two possible ways. First, there will be a bound on the sound velocity which is given by $a\omega_0$. However, considering the ratio $D_{th}/D_P$, since the sound velocity affects both $D_{th}$ and $D_P$ in the same way, it does not influence the breaking of Planckian diffusivity. Secondly, the bound on $\omega_0$ will also set an upper bound for the phonon frequency $\omega(\vec{q})$. To stay in the regime where the acoustic phonons are well-defined, the momentum-dependent relaxation time $\tau(\vec{q})$ will be bounded from below. Nevertheless, since the melting temperature should be a property of the intercell bond, such bound on lifetime should not depend on the internal property of the unit cell. As a result, the thermal diffusivity should still scale as $1/N$ as the number of intra-cell degrees of freedom increases.

The mechanism for sub-Planckian heat diffusivity here constitutes of large number of uncorrelated phonons that contributes to entropy but not heat transport. It is interesting to note that these ingredients can also show up in amorphous solids or glasses. Nevertheless, the disorder in these systems does not guarantee fast thermalization. Specifically, one can imagine the appearance of many harmonic modes in a disorder system that does not relax energy. In contrast, due to the strong nonlinearity of our system, the energy in a phonon excitation would relax rapidly to equilibrium. In the regime above the Ioffe-Regel limit, we believe the breaking of Planckian bound to be also possible in amorphous solids or glasses following the mechanism presented in our system. In fact, the possibility of reaching the Planckian bound has been discussed in Ref.~\cite{behnia2019lower}. 
However, thermal conductivity simulations of such systems require sophisticated method \cite{sheng1991heat}. The model we present here utilizes the timescale separation between intercell and intracell motion, enabling the perturbative approach. Therefore, it can be simulated in a straightforward way.

Even though the mechanism for sub-Planckian heat diffusivity here is rather simple in the sense that it arises from an extra contribution to heat capacity from optical phonons, this mechanism involves transport of heat without the presence of well-defined waves. The anharmonic nature of interactions of the balls in the shell can be viewed as a strongly interacting (although classical) phonon system which is very inefficient in carrying the stored entropy. Understanding the temperature dependence of our results would require us to go to lower temperatures where some of the higher frequency dynamics would "freeze" out as the Bose-Einstein distribution replaces the equipartition theorem. However, this regime is beyond the validity of our formalism and provides an opportunity for studying quantum chaotic dynamics. In this case, it becomes a difficult quantum many-body problem where the present approach is invalid.

JS thanks Aharon Kapitulnik and Sean Hartnoll for valuable comments. This work was supported by NSF DMR1555135 (CAREER) and JQI-NSF-PFC (supported by NSF grant PHY-1607611). This research was partially supported (through helpful discussions at KITP) by the National Science Foundation under Grant No. NSF PHY-1748958.

\appendix
\section{Thermal conductivity in terms of two-point correlation functions}
\label{appendix:A}
The Green-Kubo formula for thermal conductivity $\kappa$ is written as
\begin{align}
    \kappa = \frac{1}{k_BT^2}\int^\infty_0 d\tau\sum_n\left\langle q_{i,\alpha}(t)q_{i+n,\alpha}(t+\tau)\right\rangle
    \label{eq:A1}
\end{align}
where $q_{i,\alpha}$ is the heat flux in the $\alpha$ direction at site $i$
\begin{align}
    q_{i,\alpha} = -\frac{\vec{f}(i,\alpha)\cdot\left(\dot{\vec{r}}_i+\dot{\vec{r}}_{i+\alpha}\right)}{2}.
\end{align}
As mentioned in the main text, for the spring force $\vec{f}(i,\alpha)$, we use the linearized form,
\begin{align}
    \vec{f}(i,\alpha)=-k(\vec{r}_i-\vec{r}_{i+\alpha})-\lambda(\dot{\vec{r}}_i-\dot{\vec{r}}_{i+\alpha})+\vec{\eta}(i,\alpha)
\end{align}
Due to the isotropy, we can simply pick $\alpha=x$ without loss of generality. Furthermore, the contributions to $\kappa$ in Eq. \ref{eq:A1} from motion in the $x,y,z$ directions are the equal. This enables us to rewrite $\kappa$ as
\begin{align}
    \kappa = \frac{3}{k_BT^2}\int^{\infty}_0d\tau\sum_n\left\langle q^x_{i,x}(t)q^x_{i+n,x}(t+\tau)\right\rangle
    \label{eq:A4}
\end{align}
where 
\begin{align}
    q^x_{i,x} =&\left[ -k(r_{x,i}-r_{x,i+x})-\lambda(v_{x,i}-v_{x,i+x})+\eta^x(i,x)\right]\nonumber\\ &\times(v_{x,i}+v_{x,i+x})
\end{align}
with $v=\dot{r}$.
As discussed in App. \ref{appendix:B}, our system shown linear behavior. By Wick't theorem, we can simplify quartic terms in the average to products of two-point correlation functions. Using time-reversal symmetry, the non-vanishing terms are
\begin{align}
    \langle q^x_{i,x}(t)&q^x_{i+n,x}(t+\tau)\rangle =\nonumber\\
    k^2\{&[2C_{xx}(n,\tau)-C_{xx}(n-x,\tau)-C_{xx}(n+x,\tau)]\nonumber\\&\times[2C_{vv}(n,\tau)+C_{vv}(n-x,\tau)+C_{vv}(n+x,\tau)]\nonumber\\+&[C_{xv}(n-x,\tau)-C_{xv}(n+x,\tau)]\nonumber\\&\times[C_{vx}(n-x,\tau)-C_{vx}(n+x,\tau)]\}\nonumber\\+\lambda^2\{&[2C_{vv}(n,\tau)-C_{vv}(n-x,\tau)-C_{vv}(n+x,\tau)]\nonumber\\&\times[2C_{vv}(n,\tau)+C_{vv}(n-x,\tau)+C_{vv}(n+x,\tau)]\nonumber\\+&[C_{vv}(n-x,\tau)-C_{vv}(n+x,\tau)]\nonumber\\&\times[C_{vv}(n-x,\tau)-C_{vv}(n+x,\tau)]\}\nonumber\\ + 2\lambda& k_BT\delta_{n,0}\delta(\tau)[2C_{vv}(0,0)+C_{vv}(-x,0)+C_{vv}(x,0)].
    \label{eq:c-c correlation}
\end{align}
where $C_{AB}(n,\tau) \equiv \langle A_i(t)B_{i+n}(t+\tau)\rangle$ and the last term corresponds to the contribution from $\eta$.
Substituting Eq. \ref{eq:c-c correlation} into Eq. \ref{eq:A4} and performing a Fourier transform, the thermal conductivity can be represented as
\begin{align}
    \kappa = \frac{12}{k_BT^2}&\int_{\vec{q},\omega} \omega^2(k^2+\omega^2\lambda^2)\sin^2q_x \left[S_{xx}(\vec{q},\omega)\right]^2\nonumber\\+\frac{6\lambda}{T}&\int_{\vec{q},\omega}\omega^2(1+\cos^2q_x)S_{xx}(\vec{q},\omega)
\end{align}
where $\int_{\vec{q},\omega}\equiv\int\frac{d^3q}{(2\pi)^3}\frac{d\omega}{2\pi}$ and the relations $S_{vv} = \omega^2S_{xx}$, $S_{xv} = i\omega S_{xx}$, and $S_{vx} = -i\omega S_{xx}$ are applied.

\section{Effective Damping from Cubic Anharmonicity}
\label{appendix:B}
To work in the regime where the internal force described by $\chi$ is linear, we consider the spring force from thermal fluctuation to be weaker than the force $\phi$ from intra-cell dynamics, that is, $k\cdot k_BT < \left\langle \phi^2\right\rangle$. In the harmonic limit ($A\rightarrow 0$), the response function in momentum space defined by 
is given by
\begin{align}
    D_0(\vec{q},\omega) = \frac{1}{\chi^{-1}(\omega)+k\left(\vec{q}\right)},
\end{align}
where $\chi(\omega)$ is the Fourier transform of $\chi(t)$, $k(\vec{q}) = 4k\sum_{u\in \{x,y,z\}}\sin^2 q_u/2$ is the spring force in momentum space. Since the center of mass coordinate of each unit cell is free from the intra-cell force, there will be well-defined acoustic phonon peaks in $D_0$ at long wavelength. 

From a perturbative picture, the appearance of anharmonicity gives rise to broadening in these coherent peaks through an effective damping force,  $-\lambda\sum_{\alpha\in\{x,y,z\}}(2\dot{\vec{r}}_i-\dot{\vec{r}}_{i+\alpha}-\dot{\vec{r}}_{i-\alpha})$ on $\vec{r}_i$. At finite temperature, the anharmonic force can be thought of as a driving force on the harmonic oscillator. At a timescale larger than the correlation time of the anharmonic force $\vec{f}^A(i,\alpha)$, one can approximate the anharmonic force by a stochastic random force $\vec{\eta}(t)$ with correlation function given by
\begin{align}
    \left\langle\eta_u(t)\eta_v(t')\right\rangle=g\delta_{u,v}\delta(t-t')
\end{align}
where $g$ is the fluctuation strength that is given by the correlation function of $\vec{f}^A(i,\alpha)$,
\begin{align}
    g = \int_{-\infty}^{\infty} d\tau \left\langle f^A_u(i,\alpha,t+\tau) f^A_u(i,\alpha,t) \right\rangle_0 - \left\langle f^A_u(i,\alpha,t) \right\rangle_0^2.
    \label{eq:g}
\end{align}
where  $\left\langle \cdot\right\rangle_0$ denotes expectation values taken in the harmonic limit. Due to the isotropy, $u$ can be taken as either the $x,y$, or $z$ component of the non-linear force in Eq. \ref{eq:spring_force}.  By applying Wick's theorem, $g$ can be written in terms of the power spectrum in the following integral form. 
\begin{align}
    &g =  32A^2\int \frac{d\omega}{2\pi}\mathcal{I}(\omega)\mathcal{I}(-\omega), \nonumber\\
    &\mathcal{I}(\omega) = \int \frac{d^3q}{(2\pi)^3}\left(\sum_{u\in\{x,y,z\}}\sin^2\frac{q_u}{2}\right)  S_{0,xx}(\vec{q},\omega).
\end{align}
The effective damping $\lambda$ is then given by the following form according to fluctuation-dissipation theorem~\cite{kubo1966fluctuation}:
\begin{align}
    \lambda = \frac{g}{2k_B T}
    \label{Eq:damping}
\end{align}
Note that since $g$ is quartic in displacements (Eq.~\ref{eq:g}), according to the Wick's theorem and the equipartition principle, we expect $g$ to scale as $T^2$. Together with Eq.~\ref{Eq:damping}, the scaling of damping is $\lambda\sim~T$. In the linear response regime, including the effect of $\lambda$ into the $D_0(\vec{q},\omega)$ gives the full response function,
\begin{align}
    D(\vec{q},\omega) &= \frac{1}{D_0^{-1}(\vec{q},\omega)-4i\omega\lambda\sum_{u}\sin^2q_u/2}\nonumber\\
    &=\frac{1}{\chi^{-1}(\omega)+4(k-i\omega\lambda)\sum_{u}\sin^2 q_u/2}
\end{align}
where $u$ sums over $x,y,z$.
\section{Velocity Distribution of the Shell-ball Unit Cell}
\label{appendix:C}
\begin{figure}
    \centering
    \includegraphics[width = 0.46\textwidth]{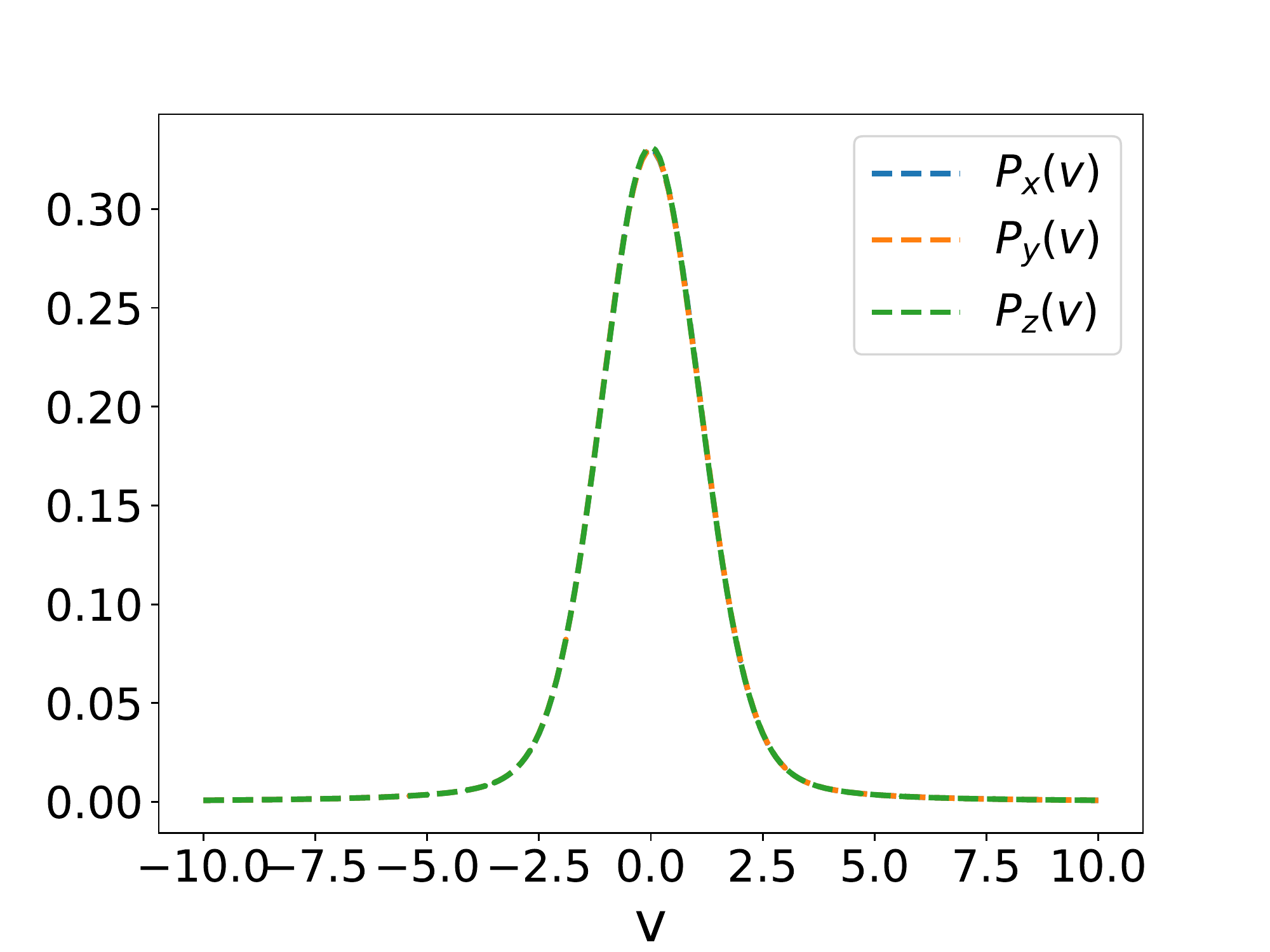}
    \caption{Velocity probability distribution of the shell coordinate in different directions for $N=40$ obtained from simulation with $5\times10^6$ collisions. The distribution functions match with each other and exhibit Gaussian shapes.}
    \label{fig:velocity_distribution}
\end{figure}
During our simulation, the shell motions between collisions are free. Therefore, within the time interval $t_{i+1}-t_i$ between the $i^{th}$ and $(i+1)^{th}$ collisions, the shell velocity $\vec{v}_i = (v_{x,i},v_{y,i},v_{z,i})$ is a constant. In this case, it is straightforward to define the velocity distribution as
\begin{align}
    P_\alpha(v) = \sum_i \frac{t_{i+1}-t_{i}}{T}\delta(v_{\alpha,i}-v)
\end{align}
where $\alpha\in\{x,y,z\}$ labels the components of the velocity and $T = t_f-t_0$ is the total time. To get a smooth probability distribution, we broaden the $\delta$ functions by Lorentzians with width $\Gamma=0.5$. The result for $N = 40$, number of collisions = $5\times10^6$ is shown in Fig. \ref{fig:velocity_distribution}. As can be seen, the velocity distributions in the $x,y$ and $z$ directions match with each other, indicating that our simulation has reached statistical equilibrium. More importantly, the Gaussian nature of $P_\alpha(v)$ validates the application of linear response theorem on the shell coordinate $\vec{r}_i$ in the main text.
\bibliography{main}

\end{document}